\documentclass[ a4paper, 12pt]{article}
\usepackage[text={18cm,24cm},centering]{geometry} % geometrija stranice

\usepackage[multiple]{footmisc}
\usepackage{appendix}

\usepackage[utf8]{inputenc}
\usepackage{amssymb ,amsmath, amsthm}
\usepackage{graphicx} % paket za umetanje slika
\usepackage{color}
\usepackage{float}
\usepackage{ulem}
\usepackage{multirow}
\usepackage{authblk}

\newtheorem{tm}{Theorem}[section]

\usepackage[font=footnotesize, labelfont=bf]{caption}
\newcounter{Section}
\newcommand{\Keywords}[1]{\par\noindent
{\small{\em Keywords\/}: #1}}

\title{Multi-scale Representation of High Frequency Market Liquidity}

\author[1]{Anton Golub\footnote{Corresponding author: agolub@olsen.ch}}
\author[2]{Gregor Chliamovitch} 
\author[1,2]{Alexandre Dupuis}
\author[2]{Bastien Chopard}
\affil[1]{\footnotesize{Olsen Ltd, av. de la Gare 8, 1870 Monthey, Switzerland}}
\affil[2]{\footnotesize{Computer Science Department, University of Geneva, rte de Drize 7, 1227 Carouge, Switzerland}}

\date{\today}

\begin{document} 

\maketitle 

\abstract{We introduce an event based framework of directional changes and overshoots to map continuous financial data into the so-called Intrinsic Network - a state based discretisation of intrinsically dissected time series. Defining a method for state contraction of Intrinsic Network, we show that it has a consistent hierarchical structure that allows for multi-scale analysis of financial data. We define an information theoretic measurement termed Liquidity that characterises the unlikeliness of price trajectories and argue that the new metric has the ability to detect and predict stress in financial markets. We show empirical examples within the Foreign Exchange market where the new measure not only quantifies liquidity but also acts as an early warning signal. \\ \Keywords{Liquidity, Information Theory, Multi-Scale, Foreign Exchange, High Frequency Trading}
}

\maketitle
\section{Introduction}

The notion of market liquidity is nowadays almost ubiquitous. It
quantifies the ability of a financial market to match buyers and sellers
in an efficient way, without causing a significant movement in the
price, thus delivering low transaction costs. It is the lifeblood of
financial markets (Fernandez 1999) without which market dislocations can
show as in the recent well documented crisis: 2007 Yen carry trade
unwind (Brunnermeier \textit{et. al.} 2008), 2008 Credit Crunch
(Brunnermeier 2009), May 6th 2010 Flash Crash (Kirilenko \textit{et.
al.} 2011, SEC 2011) or the numerous Mini Flash Crashes (Golub
\textit{et. al.} 2012, Johnson \textit{et. al.} 2013) occurring in US
equity markets, but also in many others cases that go unnoticed but are
potent candidate to become more important. While omnipresent, liquidity
is an elusive concept. Several reasons may account for this ambiguity;
some markets, such as the foreign exchange (FX) market with the daily
turnover of \$5.3 trillion (BIS 2013), are mistakenly assumed to be
extremely liquid, whereas the generated volume is equated with
liquidity. Secondly, the structure of modern markets with its high
degree of decentralization generates fragmentation and low transparency
of transactions which complicates the way to define market liquidity as
a whole. For instance, the implementation of Regulation NMS in US equity
markets has created a fragmented ecosystem where trading is split
between 13 public exchanges, more than 30 dark pools and over 200
internalizing broker-dealers (Shapiro 2010). Aggregating liquidity from
all trading sources can be quite daunting and even with all of the
market fragmentation, as new venues with different market structure
continue to be launched. Furthermore, the landscape is continuously
changing as new players emerge, such as high frequency traders that have
taken over the role of liquidity intermediation in many markets,
accounting between 50\% and 70\% of all trading (Chaboud \textit{et al.} 2012). The shift in liquidity
provision was the result of legislative changes (in US Regulation NMS,
2005 and in Europe MiFID, 2007) which was fostered greater competition
and prompted by substantial technological advances in computation and
communication that made high-speed trading possible between different
trading venues. Last, but not least, important participants influencing
the markets are the central banks with their myriad of market
interventions, whereas it is indirectly through monetization of
substantial amount of sovereign and mortgage debt with various
quantitative easing programs, or in a direct manner as with Swiss
National Bank setting the floor on EUR/CHF exchange rate, providing
plenty of arguments they have overstepped their role of last resort
liquidity providers and at this stage they hamper market
liquidity\footnote{\textit{``Minutes of the Monetary Policy Meeting on
April 26, 2013"} Bank of Japan, 2013.}\footnote{\textit{``Here Comes the
Great Bond Liquidity Crisis"}, Lee, P., September 26, 2013,
Euromoney}\footnote{\textit{``The Fed Now Owns One Third Of The Entire
US Bond Market''}, Durden, T., 2013, Zerohedge}, potentially exposing
themselves to  massive losses in the near future\footnote{\textit{``SNB
Losses 1.85 Billion Francs in Just One Day, 231 Francs per
Inhabitant''}, Dorgan, G., 2012, SNBCHF.com}\footnote{\textit{``SNB
Losses in October and November: 8.4 Billion Francs, 1.5\% of GDP''},
Dorgan, G., 2012, SNBCHF.com}\footnote{\textit{``The Fed's Impersonation
Of The Hunt Brothers Continues''}, Tchir, P., 2013, TF Market Advisors}. 

Despite the obvious importance of liquidity there is little agreement on
the best way to measure and define market liquidity (von Wyss 2004, Sarr
and Lybek 2002, Kavajecz and Odders-White 2008, Gabrielsen \textit{et
al.} 2011). Liquidity measures can be classified into different
categories. Volume-based measures: liquidity ratio, Martin index, Hui
and Heubel ratio, turnover ratio, market adjusted liquidity index (see
Gabrielsen \textit{et al.} 2011 for details) where, over a fixed period
of time, the exchanged volume is compared to price changes. This class
implies that non-trivial assumptions are made about the relation between
volume and price movements. Other classes of measures include price
based measures: Marsh and Rock ratio, variance ratio, vector
autoregressive models; transaction costs based measures: spread, implied
spread, absolute spread or relative spread see; or time based measures:
number of transactions or orders per time unit. There exists plenty of
studies that analyse these measures in various contexts (see von Wyss
2004, Gabrielsen \textit{et al.} 2011 and references therein) without
reaching a consensus. The aforementioned approaches suffer from many
drawbacks. They provide a top-down approach of analysing a complex
system, where the impact of the variation of liquidity is analysed
rather than providing a bottom-up approach where liquidity lacking times
are identified and quantified. These approaches also suffer from a
specific choice of physical time, that does not reflect the correct and
multi-scale nature of any financial market. Finally, we argue that some
of the data could be hard to get or even not available as it is the case
for the full limit order book, or trade direction, in the FX market. To
circumvent these issues and to move forward into modelling the market
dynamics, we propose an event based framework of directional changes and
overshoots to map continuous financial data into the so-called intrinsic
network - a state based discretisation of intrinsically dissected time
series, whereas the resulting structure is modelled as a multi-scale
Markov chain. We define liquidity, an information theoretic measurement
that characterises the unlikeliness of price trajectories and argue that
this new metric has the ability to detect and predict stress in
financial markets and show examples within the FX market. Finally, the
optimal choice of scales is derived using the Maximum Entropy Principle.

The rest of this paper is organised as follows; Section 2 describes the
event based framework of directional changes and overshoots. Section 3
defines the state based discretisation of price trajectory movement
termed intrinsic network. In section 4 we demonstrate the multi-scale
property of intrinsic network. In section 5 we derive the probability
matrix of the Markov chain modelling the transitions on the intrinsic
network. Section 6 describes the information-theoretic concept termed
Liquidity that characterises the unlikeliness of price trajectories. In
section 7 we discuss the optimal choice of scales applying Maximum
Entropy Principle. Finally, in section 8 we demonstrate the application
of Liquidity on 2007 Yen carry trade unwind and on Swiss National Bank
(SNB) August 2011 intervention setting the floor of 1.20 on EUR/CHF
exchange rate.

\section{The event based framework}

Traditional high frequency finance models use equidistantly spaced data
for their inputs, yet markets are known to not operate in a uniform
fashion: during the weekend the markets come to a standstill, while
unexpected news can trigger a spur of market activity. The idea of
modelling financial series using a different time clock can be traced
back to the seminal work of Mandelbrot and Taylor (1967) and Clark
(1973), advocating the use of transaction and volume based clock, moving
away from chronological time. One other area of research that analyses
high-frequency time series from the perspective of fractal theory was
initiated by Mandelbrot (1963). This seminal work has inspired others to
search for empirical patterns in market data – namely scaling
laws\footnote{A scaling law establishes a mathematical relationship
between two variables that holds true over multiple orders of
magnitude.} that would enhance the understanding of how markets work,
providing fixed points of reference and blurring out the irrelevant
details. One of the most reported scaling laws in capital markets
(Müller \textit{et al.} 1990, Galluccio \textit{et al.} 1997, Dacorogna
\textit{et al.} 2001, Di Matteo, Aste and Dacorogna 2005), relating the
average absolute price change $\langle \Delta x\rangle$ and the time
interval of its occurrence $\Delta t$: $$\langle \Delta x \rangle \sim
\Delta t^{1/2}$$ sparked an attempt to move beyond the constraints of
physical time devising a time-scale named "$\theta$-time" to account for
seasonal patterns correlated with the changing presence of main market
places in the FX market (Guillaume \textit{et al.} 1995). This approach
was not flawless, since aggregating and interpolating tick data amongst
fixed or predetermined time intervals, important information about the
market microstructure and trader behaviour is lost (Bauwens and Hautsch
2009).

\begin{figure}
\begin{center}
\includegraphics[scale=0.465]{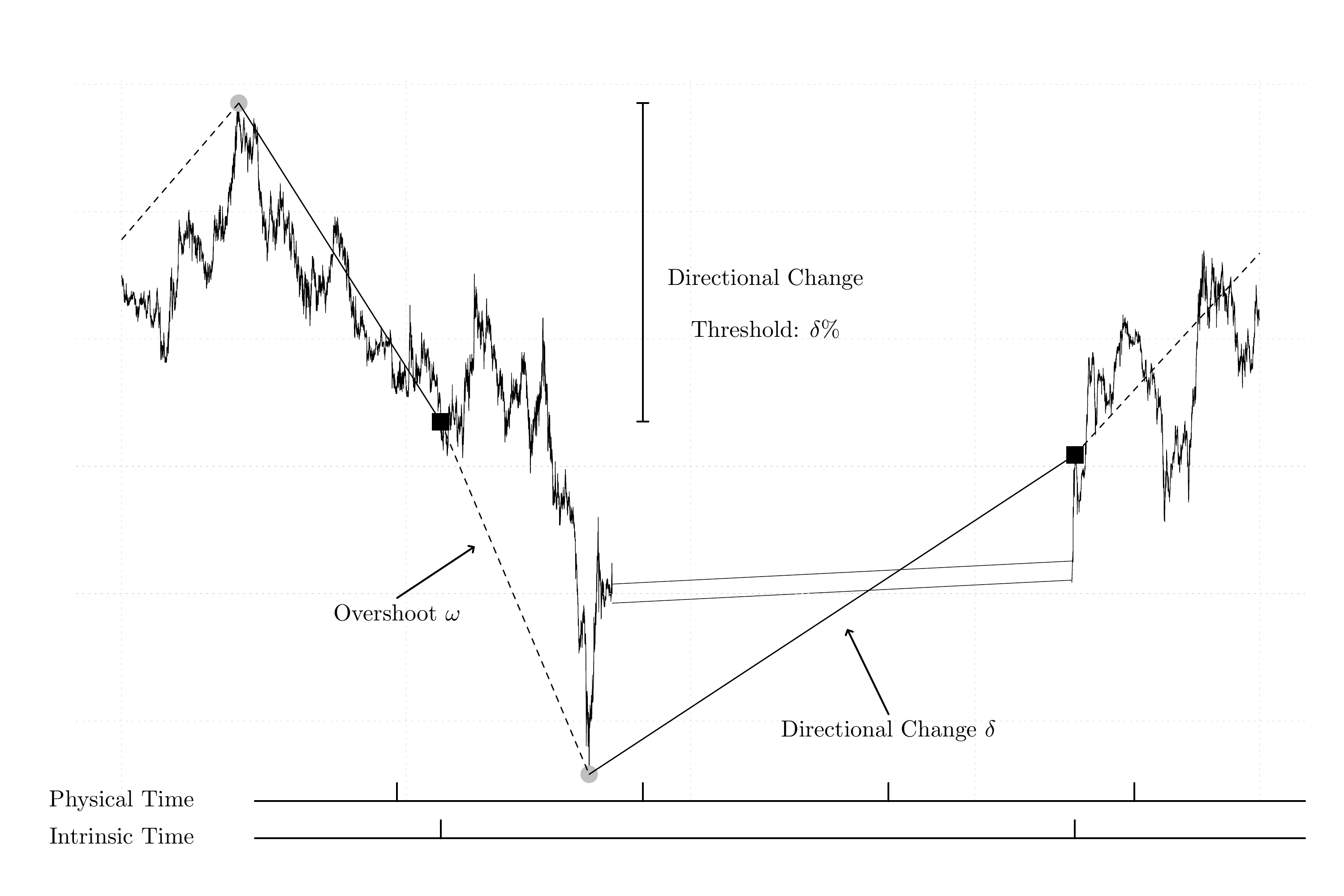}
\end{center}
\caption{Directional change events (squares) act as natural dissection points,
decomposing a total-price move between two extremal price levels
(bullets) into so-called directional-change (solid lines) and overshoot
(dashed lines) sections. Time scales depict physical time ticking evenly
across different price curve activity regimes, whereas intrinsic time
triggers only at directional change events, independent of the notion of
physical time.}
\label{fig:ccyMap}
\end{figure}

A major breakthrough occurred with the discovery of a scaling law that
relates the number of rising and falling price moves of a certain size
(threshold) with that respective threshold, which produces an
event-based time scale named ``intrinsic time'' that ticks according to
an evolution of a price move (Guillaume \textit{et al.} 1997). The
intrinsic time dissects the time series based on market events where the
direction of the trend alternates (Figure 1). These directional change
events are identified by price reversals of a given threshold value set
ex-ante. Once a directional change event has been confirmed an overshoot
event has begun and they continue the trend identified by the
directional changes. So, if an upward event has been confirmed, an
upward overshoot follows and vice versa. An overshoot event is confirmed
when the opposite directional change occurs. With each directional
change event, the ``intrinsic time'' ticks one unit. Details of the
algorithm are provided in appendix~\ref{app:algo}. The benefits of this
approach in the analysis of high-frequency data are threefold; firstly,
it can be applied to non-homogeneous time series without the need for
further data transformations. Secondly, multiple directional change
thresholds can be applied at the same time for the same tick-by-tick
data. And thirdly, it captures the level of market activity at any one
time.

Using the aforementioned framework Glattfelder et al. (2011) have
discovered 12 independent scaling laws that hold true across 13 currency
pairs across multiple orders of magnitude, linking together concepts
like the length of the overshoot, the length of the directional change,
the number of directional changes and the number of overshoot ticks with
the threshold value. Although the scaling law discoveries were
acknowledged, there was no consensus on what drives this behaviour. This
concept was predominantly used to design trading models. For instance,
in (Dupuis and Olsen 2011) and (Voicu 2012) a basic counter-trending
trading model is described which exploits profit opportunities contained
in the long coastline of prices.

Let us formalise the considerations of the framework. With $\delta >0$
we will denote the directional change threshold, as well as the
directional change itself, while the corresponding overshoot is denoted
with $\omega$. We argue that the length of the overshoot $\omega$ is a
proxy for liquidity as a long overshoot seems to imply that the market
had to further progress in the direction of the overshoot to find the
necessary liquidity to eventually retrace and exhibit the next
directional change. 
%Recall that liquidity can be viewed as the ability of the market to
%match buyers and sellers without causing a significant movement in the
%price. Therefore if the overshoot length is too long, we argue that it
%represents an illiquid market period, excreting pressure on one side of
%the market as the buying and selling interest are in imbalance.
%Likewise, if the overshoots are short we argue it represents a period
%with ample liquidity, matching buyers and sellers in an efficient
%manner. 

Claiming overshoot length is too long or too short is an abstract notion
without a reference providing an ``equilibrium'' length. The following
theorem establishes that given a stylized model of Brownian motion
driving the price movement, the expected length of the overshoot
$\omega$ equals the directional change threshold $\delta$;\newpage

\begin{tm}{(Fundamental intrinsic theorem)} 
Let the price $(P_t,t\in\mathbb{R}_+)$ be modelled as Brownian motion $(W_t,t\in\mathbb{R}_+)$ with volatility $\sigma>0$
\begin{equation}
dP_t=\sigma dW_t.
\end{equation}
Let $\omega(\delta;\sigma)$ denote the length of the overshoot for directional change threshold $\delta$. Then
\begin{equation}
\mathbb{E}[\omega(\delta;\sigma)]=\delta.
\end{equation}
\label{th:fit}
\end{tm}

The proof is provided in appendix~\ref{app:FIT}. Unexpectedly, the
average length of the overshoot $\mathbb{E}[\omega(\delta;\sigma)]$ is
completely insensitive to volatility; regardless of the volatility level
the average length of the overshoot will be proportional to the
directional change threshold. Finally, we note that the theorem
establishes a scaling law relationship between overshoot length and
directional change threshold.

\section{The intrinsic network}
\label{sec:in}

This section proposes a novel way to describe the evolution of a time
series of prices by discretisiting the price movement over various
scales and modelling it as transitions on a structure termed intrinsic
network.

We consider $n\in\mathbb{N}$ ordered thresholds
$\delta_1<\delta_2<\dots<\delta_n$ that dissect the price curve into
directional changes of fixed length $\delta_i$ and overshoots $\omega_i$
of varying length, assigning the states of the market for a given
directional change threshold $\delta_i$ either to be $1$ or $0$,
depending whether the overshoot related to the corresponding threshold
$\delta_i$ is moving upwards or downwards. With this procedure at each
physical time we can assign a binary vector $b=(b_1,\dots, b_n)$
consisting of $1$ or $0$, describing the market over various scales. The
binary encoding $b=(b_1,\dots,b_n)$ can express the state of the market
$s$ in numeric terms as follows $s=b_1\cdot 2^0+b_2\cdot 2^1 + \dots +
b_{n}\cdot 2^{n-1}$. The set of binary vectors will be denoted by
$\mathcal{B}$, set of states in numeric representation with
$\mathcal{S}$ and due to 1-1 correspondence between $\mathcal{S}$ and
$\mathcal{B}$ we will interchangeably use both notations, whereas it is
straightforward to notice that both sets have a total of $2^n$ possible
states. 

The ascending sorted thresholds and the underlying intrinsic time
dynamics imply a simple rule describing the transition between states.
Indeed in any state a move up would turn the smallest $i$ state $s_i$
showing a previous move down (i.e. $s_i=0$) into an upward state $s_i=1$
state and similarly a move down would flip the first $i$ state $s_i=1$
state into an $s_i=0$ state. The precise transition rules are stated
bellow;

\begin{center}
 state $b=(b_1,\dots,b_n)\in\mathcal{B}$ can transition to state
	$$b'=(\overline{b_1},b_2,\dots,b_n)$$ or to state
	$$b''=(b_1,b_2,\dots,\overline{b_i},\dots,b_n),~ i=\min\{k\colon b_k\neq b_1\}$$
	where $~\overline{1}=0$ and $~\overline{0}=1$.
	\label{th:transition}
\end{center}

For a given $n\in\mathbb{N}$ and directional change thresholds
$\delta_1<\dots<\delta_n$, we obtain a set of states $\mathcal{S}$ with
the transitions prescribed by the above rule. In other words, the
procedure creates a network which we call intrinsic network, denoted
with $\mathcal{IN}(n;\{\delta_1,\dots,\delta_n\};W)$ where $W$ denotes
the transition probability matrix of the stochastic process
modelling the transitions on the network.

\begin{figure}
\begin{center}
\includegraphics[width=11cm, height=13.6cm]{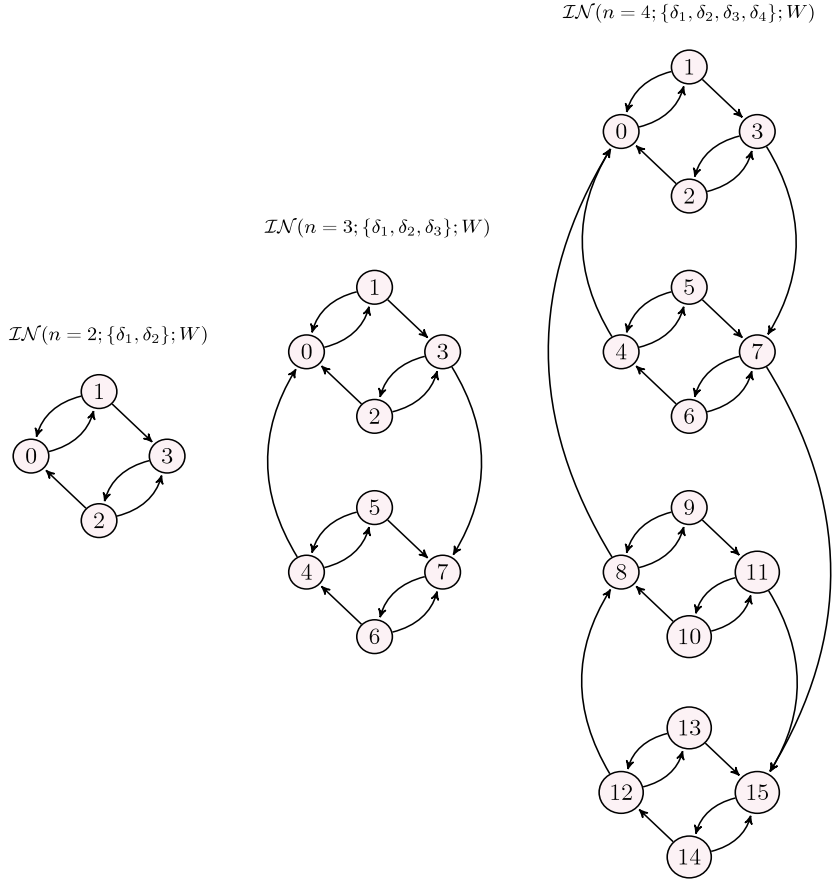}
\end{center}
\caption{From left to right: 2-, 3- and 4-dimensional intrinsic networks
$\mathcal{IN}$, with states denoted in numeric presentation.}
\label{fig:inex}
\end{figure}

The intrinsic network exhibits a couple of peculiar states where the
network is non-reactive: the downward blind-spot $(0,\dots,0)$ when the
market can keep on moving down and the upward blind-spot $(1,\dots,1)$
when the market can keep on moving up without being traced. From the
aforementioned states the intrinsic network has only one possible
transition, state $(1,1,\dots,1)$ transitions with certainty to state
$(0,1,\dots,1)$, $$\mathbb{P}\big((1,1,\dots,1)\rightarrow
(0,1,\dots,1)\big)=1,$$ while state $(0,0,\dots,0)$ transitions with
certainty to state $(1,0,\dots,0)$,
$$\mathbb{P}\big((0,0,\dots,0)\rightarrow (1,0,\dots,0)\big)=1.$$
Figure~\ref{fig:inex} shows examples of respectively 2-, 3- and
4-dimensional intrinsic networks $\mathcal{IN}$.

\section{Implicit hierarchy}
\label{sec:impHier}

In this section we demonstrate that intrinsic networks have a convenient
multi-scale property where one can dismiss the smallest threshold from
the framework and still preserve the structure. In order words, if we
remove the directional change threshold $\delta_1$ of an $n$-dimensional
intrinsic network $\mathcal{IN}(n;\{\delta_1,\dots,\delta_n\};W)$, the
resulting structure is an $n-1$-dimensional intrinsic network
$\mathcal{IN}(n-1;\{\delta_2,\dots,\delta_n\};\widehat{W})$, whereas
there is an explicit connection between transition matrices $W$ and
$\widehat{W}$, assuming the transitions on the network are modelled as
first order Markov chain process.

Firstly, we introduce the concept of islands which are subsets of all
possible states $\mathcal{S}=\{0,1,\dots,2^n-1\}$ the set of states of
an $n$-dimensional intrinsic network
$\mathcal{IN}(n;\{\delta_1,\dots,\delta_n\};W)$. We define the $k$-th
island as the following subset of states
\begin{equation}
	\mathcal{I}_k=\{s\in\mathcal{S}\colon \lfloor\frac{s}{2}\rfloor=k\},
\end{equation}
where $\lfloor \cdot \rfloor$ denotes the floor function, i.e. $\lfloor
x\rfloor =\max\{k\in\mathbb{N}\colon k\leq x\}$. In our cases, the
$k$-th island equals $\mathcal{I}_k=\{2k,2k+1\}$. For instance, island
$\mathcal{I}_0$ equals to subset $\{0,1\}$ or island $\mathcal{I}_{7}$
equals to subset $\{14,15\}$. It is easy to notice that for an
$n$-dimensional intrinsic network with $2^n$ states, there are $2^{n-1}$
islands. Note that we can again label the islands in numeric terms,
$$\mathcal{I}_k = k,$$ creating a new set of states with numeric labels
$\mathcal{S}^{(1)}=\{0,\dots,2^{n-1}-1\}$. Let us remark on the
transitions among islands $\mathcal{I}_k$. Given the numeric notation
$s=b_1\cdot 2^0+\dots+b_{n}\cdot 2^{n-1}$ for states of $\mathcal{S}$,
each state $s^{(1)}\in\mathcal{S}^{(1)}$ can be written in numeric
notation as \[s^{(1)}=b_2\cdot 2^0+\dots+b_{n}\cdot 2^{n-2},\] hence the
transitions among islands are equivalent to occurrences of directional
changes for thresholds $\delta_2,\dots,\delta_n$. In other words,
transitions among islands are insensitive to changes in the market state
related to first directional change thresholds $\delta_1$, hence the
resulting structure is an $n-1$ dimensional intrinsic network
$\mathcal{IN}(n-1;\{\delta_2,\dots,\delta_n\}, \widehat{W})$. The
probabilities of transition matrix $\widehat{W}$ and its connection to
transition matrix $W$ is established later in this section. We refer to
the presented method as state contraction.

Let us extend the concept of state contraction by defining islands of
level $0$, $\mathcal{I}_k^{(0)}$ as the states of an $n$-dimensional
intrinsic network  $$\mathcal{I}_k^{(0)}=\{k\},~k=0,1,\dots,2^n-1.$$ If
$\mathcal{S}^{(0)}=\{0,1,\dots,2^n-1\}=\{\mathcal{I}_0^{(0)},\dots,
\mathcal{I}_{2^n-1}^{(0)}\}$ denotes the states of $n$ dimensional
intrinsic network then we define islands of level $1$, in notation
$\mathcal{I}^{(1)}_k$ as subsets 
\begin{equation}
	\mathcal{I}^{(1)}_k=\{\mathcal{I}^{(0)}\in\mathcal{S}^{(0)}\colon \lfloor\frac{\mathcal{I}^{(0)}}{2}\rfloor=k\}
\end{equation}
for $k=0,1,\dots,2^{n-1}-1$. Using the aforementioned iterative process
we can obtain islands of level $j$, in notation $\mathcal{I}^{(j)}$, by
applying the process of state contraction $j$ times, hence we can then
define an intrinsic network of level $j+1$, $~\mathcal{IN}^{(j+1)}$
whereas the states space $\mathcal{S}^{(j+1)}$ are defined as islands
$\mathcal{I}^{(j+1)}$, in other words
$\mathcal{S}^{(j+1)}=\{\mathcal{I}^{(j+1)}_0,\dots,\mathcal{I}^{(j+1)}_{
2^{n-j}-1}\}$. In more general sense, we can define islands of level $i$
as a subset of islands of level $i-1$, i.e.
$$\mathcal{I}_l^{(i)}=\{\mathcal{I}^{(i-1)}\in\mathcal{S}^{(i-1)}\colon
\lfloor\frac{\mathcal{I}^{(i-1)}}{2}\rfloor =l\}.$$ Having started with
an intrinsic network with a total of $2^n$, the intrinsic network of
level $j+1$, $\mathcal{IN}^{(j+1)}$ will have a total of $2^{n-j}$
states.

\begin{figure}
\begin{center}
  \includegraphics[scale=0.4]{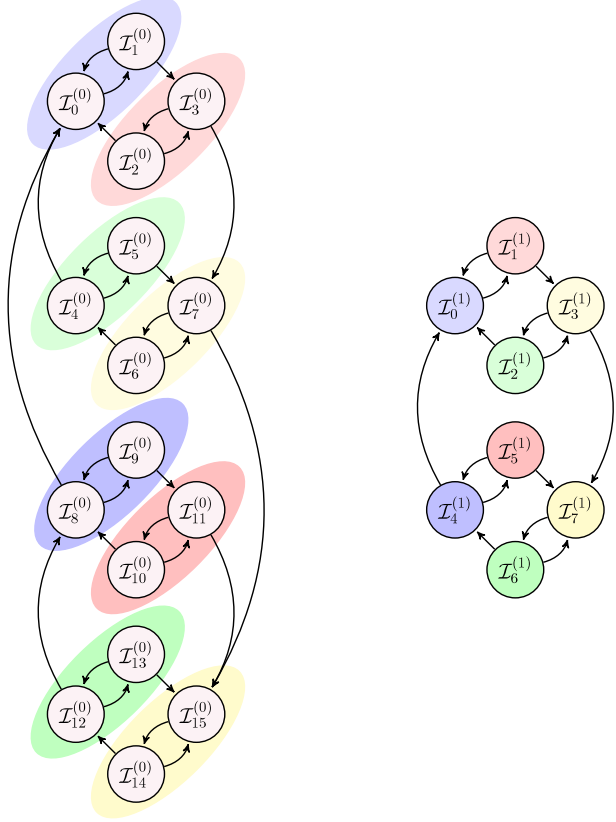}
\end{center}
\caption{Illustration of the contraction procedure of a 4-dimensional
intrinsic network
$\mathcal{IN}(4;\{\delta_1,\delta_2,\delta_3,\delta_4\};W)$ to a
3-dimensional intrinsic network
$\mathcal{IN}(3;\{\delta_2,\delta_3,\delta_4\};\widehat{W})$, whereas
coloured shading graphs islands, contracted states and the resulting new
states.}
\label{fig:hierarchy}
\end{figure}

Figure~\ref{fig:hierarchy} illustrates the process of state contraction,
of 4-dimensional intrinsic network
$\mathcal{IN}(4;\{\delta_1,\dots,\delta_4\};W)$ whereas the islands
$\mathcal{I}_0^{(0)},\dots,\mathcal{I}_{15}^{(0)}$ are contracted in the
following manner $$\mathcal{I}_i^{(1)}=\{\mathcal{I}_{2i}^{(0)},
\mathcal{I}_{2i+1}^{(0)}\}.$$ The coloured shading graphs islands,
contracted states and the resulting new states of the three dimensional
intrinsic network
$\mathcal{IN}(3;\{\delta_2,\delta_3,\delta_4\};\widehat{W})$.\\

Assuming the transitions on the intrinsic network are modelled as a
first order Markov chain, there is an explicit connection between the
transition matrix $W$ of the $n$-dimensional intrinsic network and
transition matrix $\widehat{W}$ of the $n-1$ dimensional intrinsic
network obtained through the contraction process described above. 

Let us assume that the current state of the network is
$\mathcal{I}_k^{(i)}$ and we are interested in finding the probability
of transitioning to state $\mathcal{I}_j^{(i)}$. Note that observing the
process from viewpoint of islands of level $(i-1)$, the system can
oscillate among states $\mathcal{I}_{2k}^{(i-1)}$ and
$\mathcal{I}_{2k+1}^{(i-1)}$ arbitrary many times before taking the
transition that links the islands $\mathcal{I}_k^{(i)}$ and
$\mathcal{I}_j^{(i)}$. Hence, the desired probability can be easily
derived using the closed form of geometric series. The three distinct
cases are presented bellow, with the proof in appendix~\ref{app:contrProb},

\noindent for $\operatorname{mod}(k+j,2)=0$
\begin{equation}
\mathbb{P}(\mathcal{I}_k^{(i)}\rightarrow \mathcal{I}_j^{(i)})=\frac{\mathbb{P}(\mathcal{I}^{(i-1)}_{2k+1}\rightarrow \mathcal{I}^{(i-1)}_{2j+1})}{1-\mathbb{P}(\mathcal{I}^{(i-1)}_{2k+1}\rightarrow \mathcal{I}^{(i-1)}_{2k})\cdot \mathbb{P}(\mathcal{I}^{(i-1)}_{2k}\rightarrow \mathcal{I}^{(i-1)}_{2k+1})}
\end{equation}
for $\operatorname{mod}(k+j,2)=1$ and $k>j$ 
\begin{equation}
\mathbb{P}(\mathcal{I}_k^{(i)}\rightarrow \mathcal{I}_j^{(i)})=
\frac{\mathbb{P}(\mathcal{I}^{(i-1)}_{2k}\rightarrow \mathcal{I}^{(i-1)}_{2j})\cdot \mathbb{P}(\mathcal{I}^{(i-1)}_{2k+1}\rightarrow \mathcal{I}^{(i-1)}_{2k})}{1-\mathbb{P}(\mathcal{I}^{(i-1)}_{2k}\rightarrow \mathcal{I}^{(i-1)}_{2k+1})\cdot \mathbb{P}(\mathcal{I}^{(i-1)}_{2k+1}\rightarrow \mathcal{I}^{(i-1)}_{2k})}
\end{equation}
for $\operatorname{mod}(k+j,2)=1$ and $k<j$ 
\begin{equation}
\mathbb{P}(\mathcal{I}_k^{(i)}\rightarrow \mathcal{I}_j^{(i)})=
\frac{\mathbb{P}(\mathcal{I}^{(i-1)}_{2k+1}\rightarrow \mathcal{I}^{(i-1)}_{2j+1})\cdot \mathbb{P}(\mathcal{I}^{(i-1)}_{2k}\rightarrow \mathcal{I}^{(i-1)}_{2k+1})}{1-\mathbb{P}(\mathcal{I}^{(i-1)}_{2k}\rightarrow \mathcal{I}^{(i-1)}_{2k+1})\cdot \mathbb{P}(\mathcal{I}^{(i-1)}_{2k+1}\rightarrow \mathcal{I}^{(i-1)}_{2k})}
\end{equation}

\section{Transition probabilities}

In this section we present the transition probabilities of an intrinsic
network modelling the transitions as a first order Markov chain
process.

Firstly, we stress that given a Brownian motion $(\sigma
W_t,t\in\mathbb{R}_+)$ as the process modelling the price movement
$(P_t,t\in\mathbb{R}_+)$
\begin{equation}
dP_t=\sigma dW_t,
\end{equation}
the resulting stochastic process of transitions on intrinsic networks,
in notation $(X_{\tau_{\alpha}})$, where $(\tau_{\alpha}, \alpha\in \mathcal{A})$
is the set of intrinsic times when a transition occurred is in fact a
non-Markovian process, i.e.
\[\mathbb{P}(X_{\tau_n}=s_i|X_{\tau_{n-1}}=s_j,\dots,X_{\tau_{n-k}}=s_r)
\neq
\mathbb{P}(X_{\tau_n}=s_i|X_{\tau_{n-1}}=s_j,\dots,X_{\tau_{n-m}}=s_l),~
\forall k<m.\]
In other words, the process requires the full history to derive the
transition probabilities. The non-Markovian property steams from the
fact that directional change thresholds can have different reference
points in their memory. In what follows, for the sake of simplicity, we
will nevertheless adopt a Markovian description. Properly quantifying
the impact of such a choice is an interesting and ambitious question
that we will address in the future. However, we have noticed that, in
our case, transitions probabilities between two states happen to be
close whatever the length of the memory considered in the description,
and also that, for the application we have in mind -namely quantifying
market liquidity- our liquidity measure seems rather insensitive to
whether or not we take memory effects into account.

In order to simplify the framework we model the transitions on the
intrinsic network as a first order Markov chain process, assuming that
the transitions for threshold $\delta_1$ correspond to a length of the
overshoot of $\delta_2-\delta_1$, i.e. let $$i=\min\{k\colon b_k\neq
b_1\},$$ for $i=2$
\begin{equation}
{P}\big((b_1,b_2,\dots,b_{n})\rightarrow (b_1,\overline{b_2},\dots,b_{n})\big)=\mathbb{P}\big(\omega(\delta_1;\sigma)\geq \delta_2-\delta_1\big)
\end{equation}
\begin{equation}
{P}\big((b_1,b_2,\dots,b_{n})\rightarrow (\overline{b_1},b_2,\dots,b_{n})\big)=\mathbb{P}\big(\omega(\delta_1;\sigma)< \delta_2-\delta_1\big)
\end{equation}
where $~\overline{1}=0$ and $~\overline{0}=1$. Now we present the analytic expression for transition probabilities assuming (9)-(10);\\
\begin{tm}
Let $n\in\mathbb{N}$ and $\delta_1<\dots<\delta_n$ directional change
thresholds of intrinsic network
$\mathcal{IN}(n;\{\delta_1,\dots,\delta_n\};W)$ and $(b_1,\dots,b_n)
\in\mathcal{B}$ the current state of the market. Let $$i=\min\{k\colon
b_k\neq b_1\},$$ for $i=2$ 
\begin{equation}
\mathbb{P}\big((b_1,b_2,\dots,b_n)\rightarrow (b_1,\overline{b_2},\dots,b_n)\big)=e^{-\frac{\delta_2-\delta_1}{\delta_1}}
\end{equation}
\begin{equation}
\mathbb{P}\big((b_1,b_2,\dots,b_n)\rightarrow (\overline{b_1},b_2,\dots,b_n)\big)=1-e^{-\frac{\delta_2-\delta_1}{\delta_1}}
\end{equation} 
while for $i>2$ 
\begin{equation}
\mathbb{P}\big((b_1,b_2,\dots,b_n)\rightarrow (b_1,\dots,\overline{b_i},\dots,b_n)\big)=\frac{\prod_{k=2}^i e^{-\frac{\delta_k-\delta_{k-1}}{\delta_{k-1}}}}{1-\sum_{k=2}^{i-1}\big(1-e^{-\frac{\delta_k-\delta_{k-1}}{\delta_{k-1}}}\big)\prod_{j=k+1}^i e^{-\frac{\delta_j-\delta_{j-1}}{\delta_{j-1}}}}
\end{equation} 
\begin{equation}
\mathbb{P}\big((b_1,b_2,\dots,b_n)\rightarrow (\overline{b_1},\dots,b_i,\dots,b_n)\big)=1-\frac{\prod_{k=2}^i e^{-\frac{\delta_k-\delta_{k-1}}{\delta_{k-1}}}}{1-\sum_{k=2}^{i-1}\big(1-e^{-\frac{\delta_k-\delta_{k-1}}{\delta_{k-1}}}\big)\prod_{j=k+1}^i e^{-\frac{\delta_j-\delta_{j-1}}{\delta_{j-1}}}}
\end{equation}
where $~\overline{1}=0$ and $~\overline{0}=1$.
\end{tm}

The proof for the analytical expressions in theorem 5.1 can be found in
appendix~\ref{app:contrProb}, and it boils down to deriving the
expressions (11) and (12) and then applying the explicit formulas for
transition probabilities of contracted intrinsic network presented in
section~\ref{sec:impHier}.

\section{Price trajectory unlikeliness}

Here we introduce the Liquidity, an information theoretic value that
measures the unlikeliness of price trajectories mapped to the intrinsic
network. Before proceeding let us simplify the notation used so far. The
stochastic process modelling the transition on the intrinsic network
$\mathcal{IN}$ was denoted with $(X_{\tau_\alpha})$ where
$(\tau_{\alpha},\alpha\in\mathcal{A})$ are the times of occurrences of
transitions on the network. In other words, $(\tau_{\alpha})$ is the
waiting times process when any of the directional changes related to
thresholds $\delta_i,i=1,\dots,n$ has occurred. Hence, modelling the
transitions on the network as a first order Markov chain, the
corresponding probabilities are denoted with
\[\mathbb{P}(X_{\tau_n}=s_i|X_{\tau_{n-1}}=s_j),~s_i,s_j\in\mathcal{S}.\] 
We will shorten this notation by writing only that a transition is made
from state $s_i$ to $s_j$, neglecting the intrinsic times when the
transition occurred, i.e. we will simplify the notation by writing
\[\mathbb{P}(s_i\rightarrow s_j).\]

Firstly, let $\mathcal{IN}(n;\{\delta_1,\dots,\delta_n\};W)$ denote the
$n$ dimensional intrinsic network with ordered directional change
thresholds $\delta_1<\dots<\delta_n$, state space
$\mathcal{S}=\{0,1,\dots,2^n-1\}$ and let $W$ denote the corresponding
transition probability matrix $$W=\big[\mathbb{P}(s_i\rightarrow
s_j)\big]_{i,j=0}^{2^n-1}.$$ We define the surprise of transition from
state $s_i$ to state $s_j$, in notation $\gamma_{s_i,s_j}$, by the
following expression
\begin{equation}
\gamma_{s_i,s_j}=-\operatorname{log}\mathbb{P}(s_i\rightarrow s_j).
\end{equation}

Let us explain the intuition behind the surprise $\gamma_{s_i,s_j}$, if
the transition from state $s_i$ to $s_j$ is very unlikely, i.e.
$\mathbb{P}(s_i\rightarrow s_j)\approx 0$, the corresponding surprise
$\gamma_{s_i,s_j}$ will be very large, i.e. $\gamma_{s_i,s_j}\gg 0$,
implying that the price trajectory experienced an unlikely movement
within the context of its intrinsic network. Note that the
above-introduced surprise of transition should be familiar to anyone
working in information theory, since entropy is obtained by simply
averaging the surprise. In other words, entropy is the average
uncertainty of the process, while surprise is this same quantity
evaluated for a particular realization of the process, i.e. the surprise
is a point-wise entropy.

Let us now assume that we are observing the price trajectory $(P_t)$
within a time interval $[0,T],~T>0$ and within this time interval the
price trajectory experienced $K$ transitions $$s_{i_1}\rightarrow
s_{i_2}\rightarrow \dots \rightarrow s_{i_K}\rightarrow s_{i_{K+1}}$$
hence residing in $K+1$ states during time interval $[0,T]$. We define
the surprise of price trajectory within time interval $[0,T]$, in
notation $\gamma^{[0,T]}_{s_{i_1},\dots,s_{i_{K+1}}}$ as \begin{align}
\gamma^{[0,T]}_{s_{i_1},\dots,s_{i_{K+1}}}&=-\operatorname{log}\mathbb{P
}(s_{i_1}\rightarrow s_{i_2}\rightarrow s_{i_3}\rightarrow \dots
s_{i_K}\rightarrow s_{i_{K+1}})\\
&=\sum_{k=1}^K-\operatorname{log}\mathbb{P}(s_{i_k}\rightarrow
s_{i_{k+1}})\\ &=\sum_{k=1}^K \gamma_{s_{i_k},s_{i_{k+1}}} \end{align}

If the $K$ transitions taken on the intrinsic network are clear from the
context, we will denote the surprise with $\gamma^{[0,T]}_K$. The
defined value, as sum of unlikeliness of individual transitions,
measures the unlikeliness of price trajectory within a given time
interval. Should the value be very large, it would indicate that the
trajectory experienced an unlikely movement. Furthermore, it is
important to stress that the surprise is a price trajectory dependent
measurement: two price trajectories of same volatility $\sigma$ can have
widely different surprise values.

Suppose one wants to compute the surprise of a price trajectory within a
certain time interval $[0,T]$ using empirical data, applying the formula
listed above. An easily noticeable caveat arises as the price movement
can result in a few transitions on the intrinsic network in some time
periods, while others time periods can be marked by higher activity.
Therefore, by construction the hectic time periods will have higher
surprise purely due to higher number of transitions. In order to remove
the market activity from the analysis, we need to remove the component
related to the number of transitions from the surprise within the
mentioned time interval. From Shannon-McMillan-Brieman theorem on
convergence of sample entropy and the corresponding Central Limit
Theorem (Pfister \textit{et. al.} 2001), we notice that the centring
value in the expression is in fact equal to $K\cdot H^{(1)}$, while the
rescaling value equals $\sqrt{K\cdot H^{(2)}}$, involving the first
order informativeness (i.e. entropy)
\[H^{(1)}=\sum_{i=0}^{2^n-1}\mu_i \mathbb{E}[-\operatorname{log}\mathbb{P}(s_i\rightarrow \cdot)],\] 
and second orders of informativeness
\[~H^{(2)}=
\sum_{i=0}^{2^n-1}\sum_{j=0}^{2^n-1}\mu_i\mu_j\operatorname{Cov}\big(
-\operatorname{log}\mathbb{P}(s_i\rightarrow\cdot),-\operatorname{log}
\mathbb{P}(s_j\rightarrow)\big).\]
where $\mu$ is the stationary distribution of the corresponding Markov
chain, i.e. left normalised eigenvector of transition matrix $W$. The centred and rescaled
expression converges to normal distribution(Pfister \textit{et.
al.} 2001)
\begin{equation}
\displaystyle\frac{\gamma^{[0,T]}_{K}-K\cdot H^{(1)}}{\sqrt{K\cdot H^{(2)}}}\rightarrow N(0,1)~~\text{as}~~ K\rightarrow \infty,
\end{equation}
hence we call Liquidity\footnote{When Shannon had invented his quantity
and consulted von Neumann how to call it, von Neumann replied:
\textit{``Call it entropy. It is already in use under that name and
besides, it will give you a great edge in debates because nobody knows
what entropy is anyway''}, (Denbigh 1981). Using the same rational we
decided to call our measurement Liquidity.} the corresponding quantile
\begin{equation}
\mathcal{L}=\displaystyle 1-\Phi\bigg(\frac{\gamma^{[0,T]}_K-K\cdot H^{(1)}}{\sqrt{K\cdot H^{(2)}}}\bigg),~K\gg 0
\end{equation}
where $\Phi$ is the cumulative distribution function of normal
distributions. The Liquidity $\mathcal{L}$ provides an activity-free
measurement of price trajectory unlikelines where value close to zero
indicates illiquid market conditions as overshoots, observed in a
multi-scale framework, are long; while values close to one indicates
ample liquidity in the market and overshoots are short. Note that $K$,
the number of transitions within time interval $[0,T]$ on the intrinsic
network, which is explicitly expressed in the Liquidity formula, in fact
equals to the sum of all directional changes related to thresholds
$\delta_1,\dots,\delta_n$ that occurred within time interval $[0,T]$,
representing the measurement of activity across multiple scales. Hence
by subtracting $K\cdot H^{(1)}$ from surprise $\gamma^{[0,T]}_{K}$ and
rescaling it by $\sqrt{K\cdot H^{(2)}}$ we remove the contribution of
market activity from the analysis of price trajectory, allowing us to
purely observe the lengths of the overshoots in a multi-scale framework.

\section{Preferred Scales}

Throughout the paper we have been elusive about optimal setting of the
thresholds that map the continuous financial data into the intrinsic
network. In this section we discuss the optimal choice of directional
change thresholds, exploring three possibilities;
\begin{itemize}
	\item setting all achievable transition probabilities to be equally likely, i.e.  $\mathbb{P}(s_i\rightarrow s_j)=0.5,~ \newline\{s_i\rightarrow s_j\}\neq \emptyset, s_i,s_j\in\mathcal{S},$
	\item maximizing the entropy, i.e. the first order informativeness $H^{(1)}$, of the Markov chain process modelling the transitions on the intrinsic network,
	\item  maximizing the second order informativeness, $H^{(2)}$ of the Markov Chain process and relating the process to optimal setting of thresholds for Liquidity measurement $\mathcal{L}$.
\end{itemize}

Notice that all of the mentioned optimization methods can be viewed as
the consequence of application of maximum entropy principle on the
selected probability distributions in the framework. In a nutshell, the
maximum entropy principle states (Jaynes 1957a, Jaynes 1957b) that

\begin{center}
\textit{''...subject to precisely stated prior data (such as a proposition that expresses testable information), the probability distribution which best represents the current state of knowledge is the one with largest entropy.''}
\end{center}

First we discuss the resulting thresholds when setting all of the
achievable transition probabilities to be equally likely. Since the
transition from each state $s_i\in\mathcal{S}$ can be viewed as
Bernoulli random variable, its maximal entropy is achieved by setting
the probability of all non-trivial transitions to be equally likely,
i.e. $$\mathbb{P}(s_i\rightarrow s_i^1)=\mathbb{P}(s_i\rightarrow
s_i^2)=0.5~~\Rightarrow~~
\mathbb{E}\big[-\operatorname{log}\mathbb{P}(s_i\rightarrow
\cdot)\big]=\operatorname{log}2,~\forall i=1,\dots,2^n-2.$$ In that case
we can directly apply the analytical formulas for probabilities of
transitions presented in section 4, and given the first thresholds
$\delta_1$, the $k$-th threshold equals
\begin{equation}
\delta_k=\delta_1\prod_{i=1}^k\big(1+\operatorname{log}(1+\frac{1}{k}) \big).
\label{eq:delta1}
\end{equation}

We observe that such a choice of transition probabilities
induces approximately linear setting of thresholds as the limit of the
product approaches fast its limit \[\lim_{k\rightarrow
\infty}\frac{\prod_{i=1}^k\big(1+\operatorname{log}(1+\frac{1}{k})}{k}=0
.8625576..., \] in other words, for $k\gg 0$, the $k$-th directional
change thresholds $\delta_k$ is approximately \[\delta_k\approx
\delta_1\cdot k\cdot 0.8625576.\] Unfortunately, setting the thresholds
as in (\ref{eq:delta1}) disables the use of surprise
$\gamma_{K}^{[0,T]}$ in studying Liquidity as it becomes completely
deterministic and proportional to number of transitions $K$
\[\gamma_K^{[0,T]}\sim K\cdot \operatorname{log}2.\]

Secondly, the potential method for setting the thresholds would be to
maximise the entropy of the corresponding Markov chain modelling the
transitions on the intrinsic network,
\[(\delta_1^*,\dots,\delta_n^*)=\underset{(\delta_1,\dots,\delta_n)}{
\operatorname{arg~max}}~H^{(1)}=\underset{(\delta_1,\dots,\delta_n)}{
\operatorname{arg~max}}~\sum_{i=0}^{2^n-1}\mu_i
\mathbb{E}[-\operatorname{log}\mathbb{P}(s_i\rightarrow \cdot)],\] where
$\mu$ is the corresponding stationary distribution, i.e. left normalized
eigenvector of $W$. 

For a two dimensional intrinsic network
$\mathcal{IN}(2;\{\delta_1,\delta_2\};W)$ setting all transition
probabilities to be equally likely and maximising the entropy $H^{(1)}$ from formula (21)
yield equivalent optimal thresholds $(\delta_1^*, \delta_2^*)$ where
\[\delta_2^*=(1+\operatorname{log}2)\cdot \delta_1^*\] for a given
$\delta_1^*$. For $n>2$ the claim does not hold.

The final proposal to set the thresholds in an optimal manner steams
from the central measurement in our analysis, the surprise
$\gamma_K^{[0,T]}$ which is known, when properly adjusted, to converge
to normal distribution for $K\gg 0$. Reshuffling the expression of
surprise, for large but fixed $K$ the distribution is approximately
normal \[\gamma_K^{[0,T]}\sim N(K\cdot H^{(1)},K\cdot H^{(2)}).\]
Normally distributed random variable $X$ with mean $\mu$ and standard
deviation $\sigma$, i.e. $X\sim N(\mu, \sigma^2)$, has entropy given by
\[H(X)=\frac{1}{2}\operatorname{log}(2\pi e\sigma^2)\] therefore entropy
of surprise equals
\[H(\gamma_K^{[0,T]})=\frac{1}{2}\operatorname{log}\big(2\pi e(K\cdot
H^{(2)})\big)\] and applying the maximum entropy principle on the
distribution of surprise $\gamma_K^{[0,T]}$ we conclude that the optimal
choice of thresholds $(\delta_1^*,\dots,\delta_n^*)$ is the one that
maximizes the second order informativeness $H^{(2)}$, i.e. \begin{align}
(\delta_1^*,\dots,\delta_n^*)&=\underset{(\delta_1,\dots,\delta_n)}{
\operatorname{arg~max}}~H^{(2)}\\ &=\underset{(\delta_1,\dots,\delta_n)}
{\operatorname{arg~max}}~\sum_{i=0}^{2^n-1}\sum_{j=0}^{2^n-1}\mu_i\mu_j
\operatorname{Cov}\big(
-\operatorname{log}\mathbb{P}(s_i\rightarrow\cdot),-\operatorname{log}
\mathbb{P}(s_j\rightarrow)\big). \label{eq:maxH2} \end{align} We note
that the aforementioned optimisation process is a complex mathematical
problem, which is intended to be solved using numerical procedures. For
small $n$, one can explore the space of solutions of (\ref{eq:maxH2})
and estimate the necessary joint probabilities. For larger $n$ an
exploration becomes computationally undoable and we suggest to set
$\delta_1$ to a fixed value and let $\delta_i=\lambda \delta_{i-1},\lambda >0$ for $i=2\dots n$.

Choosing to maximise the first or second level of informativeness, we
stress that there exists a solution for every $n\in\mathbb{N}$ as we
note that the entropy of an $n$-dimensional intrinsic network, in
notation $H^{(1)}_n$, is bounded \begin{equation}\displaystyle
H^{(1)}_n=\frac{H^{(1)}_n}{\sum_{i=0}^{2^n-1}\mu_i}=\frac{\sum_{i=0}^{2^
n-1}\mu_i \mathbb{E}[-\operatorname{log}\mathbb{P}(s_i\rightarrow
\cdot)]}{\sum_{i=0}^{2^n-1}\mu_i}\leq
\underset{s_i\in\mathcal{S}}{\operatorname{max}}\{\mathbb{E}[-
\operatorname{log}\mathbb{P}(s_i\rightarrow \cdot)]\}\leq
\operatorname{log}2 \end{equation} and it follows from the fact that
weighted arithmetic average is always less than the maximum value of the
components, as is also the second order informativeness $H_n^{(2)}$
\begin{equation} H_n^{(2)}\leq
\underset{s_i\in\mathcal{S}}{\operatorname{max}}\Big\{\operatorname{Var}
\big(-\operatorname{log}\mathbb{P}(s_i\rightarrow \cdot)\big)^2\Big\}^2.
\end{equation} 

\section{Liquidity Shocks}

In this section we illustrate the application of the proposed Liquidity
measurement $\mathcal{L}$ on well-known FX market crisis and argue that
the Liquidity measurement can be used as an early warning signal for
stress in financial markets, focusing on the August 2007 Yen carry trade
collapse and the Swiss National Bank implementation of 1.20 floor for
EUR/CHF exchange rate.

Before we proceed, let us explain the details of the intrinsic network
used in the examples presented below. Firstly, as we are concerned with
high frequency market conditions we choose the first threshold
$\delta_1$ to be $0.025\%$ ($\sim 2.5$ pips) and taking each next
thresholds as the double of its predecessor. We use a total of twelve
thresholds, i.e. \[\delta_i=2\cdot
\delta_{i-1}=2^{i-1}\cdot\delta_1=2^{i-1}\cdot 0.025\%,~~i=2,\dots,12.\]
which produces a twelve dimensional intrinsic network. Assigning the
thresholds in the mentioned manner assures that according to empirical
scaling laws (Glattfelder \textit{et al.} 2011), on average we have a
transition on the intrinsic network approximately every 5
seconds\footnote{We note that the intra-day transition occurrences are
highly non-uniform as during news announcements there can be several
transitions within one second.}. The corresponding probability
transition matrix $W$ is obtained from analytic expressions in theorem
5.1, hence providing us with all the tools to construct the intrinsic
network $\mathcal{IN}(12;\{\delta_1,\dots,\delta_{12}\};W)$ which maps
the exchange rates to state based discretisation. Next, we numerically
approximate the first $H_{12}^{(1)}=0.4604$ and second order informativeness
$H_{12}^{(2)}=0.70818$, yielding the
Liquidity measurement
\[\mathcal{L}=1-\Phi\bigg(\frac{\gamma^{[0,T]}_K-K\cdot
H_{12}^{(1)}}{\sqrt{K\cdot H_{12}^{(2)}}}\bigg).\] whereas the sliding
window is set to one day, i.e. $T=1\text{ day}$, $K$ is the number of
transitions within the sliding window. The Liquidity $\mathcal{L}$ is
graphed every minute, dismissing the inactive periods during the
week-ends. \\

First we present the Liquidity measurement $\mathcal{L}$ during the 2007
Yen carry trade unwind. The term carry trade in context of FX markets
refers to strategy of shorting low-yielding currencies and buying
high-yielding currencies, earning the interest differential. Carry
trades are usually done with a lot of leverage\footnote{By early 2007,
it was estimated that some US\$1 trillion have been staked on the yen
carry trade (Lee 2008).}, so a small movement in exchange rates can
result in huge losses, causing massive price reversals (Brunnermeier
\textit{et al.} 2008). The August 2007 USD/JPY price drop was the result
of the unwinding of large Yen carry-trade positions; many hedge funds
and banks with proprietary trading desks had large positions at risk and
decided to buy back yen to pay back low-interest loans (Chaboud
\textit{et al.} 2012). 

\begin{figure}
\begin{center}
\includegraphics[scale=0.5]{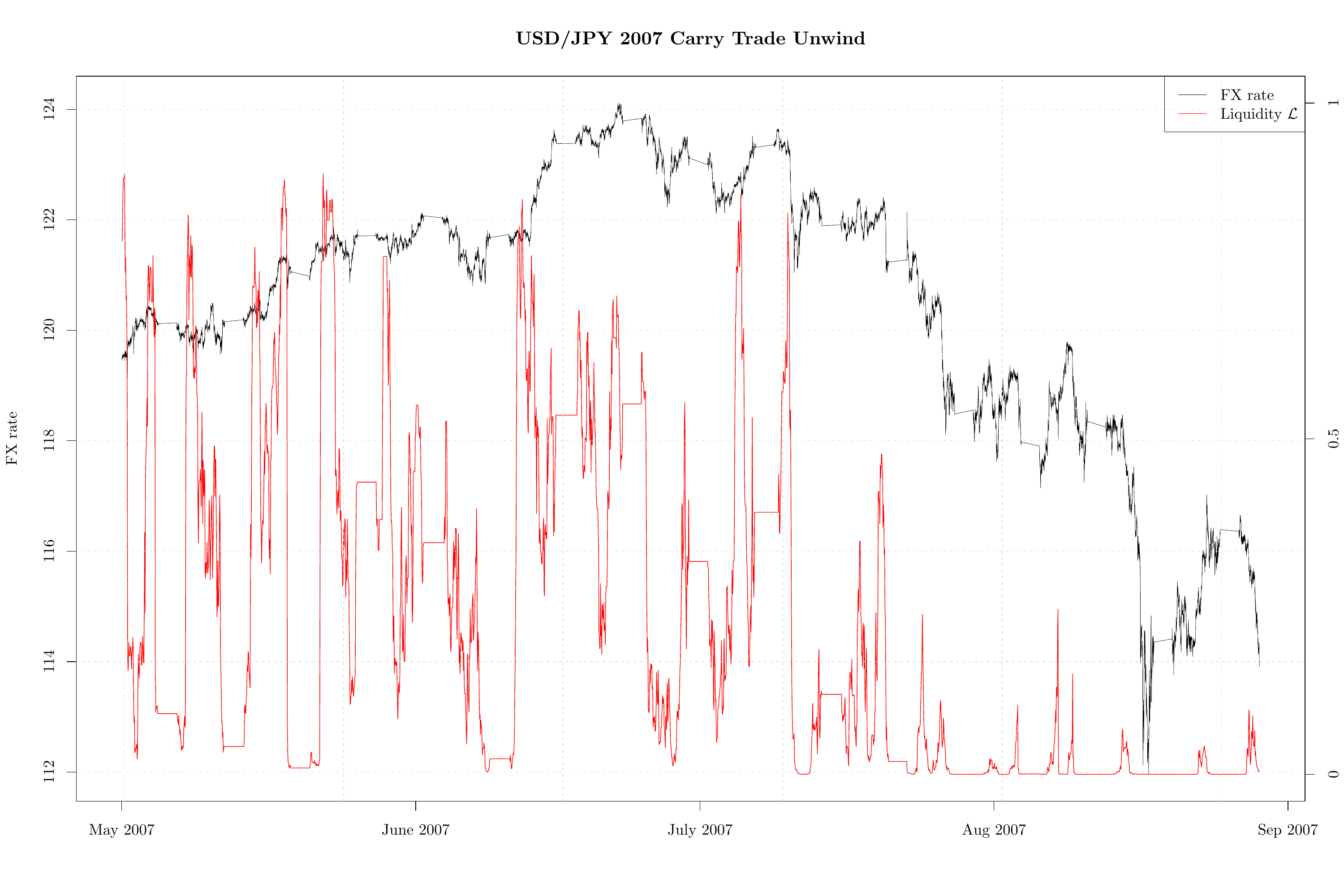}
\end{center}
\caption{
Time evolution of (left axis) tick-by-tick USD/JPY exchange rate and
(right axis) the corresponding one minute Liquidity measurement
$\mathcal{L}$ around the period of August 2007 carry trade
unwind.}
\label{fig:carry}
\end{figure}

Figure~\ref{fig:carry} shows the time evolution of the tick-by-tick
USD/JPY exchange rate and minute-by-minute Liquidity $\mathcal{L}$.
Notable shocks to market liquidity occurred in mid July with almost 2\%
drop in matter of hours. From there on, relatively illiquid conditions
is showed by the measurement, as traders started unwinding carry trade
positions. Complete loss of market liquidity is demonstrated in the
three weeks preceding the spectacular 6\% drop, which occurred on
August 16th 2007.\\

\begin{figure}
\begin{center}
\includegraphics[scale=0.5]{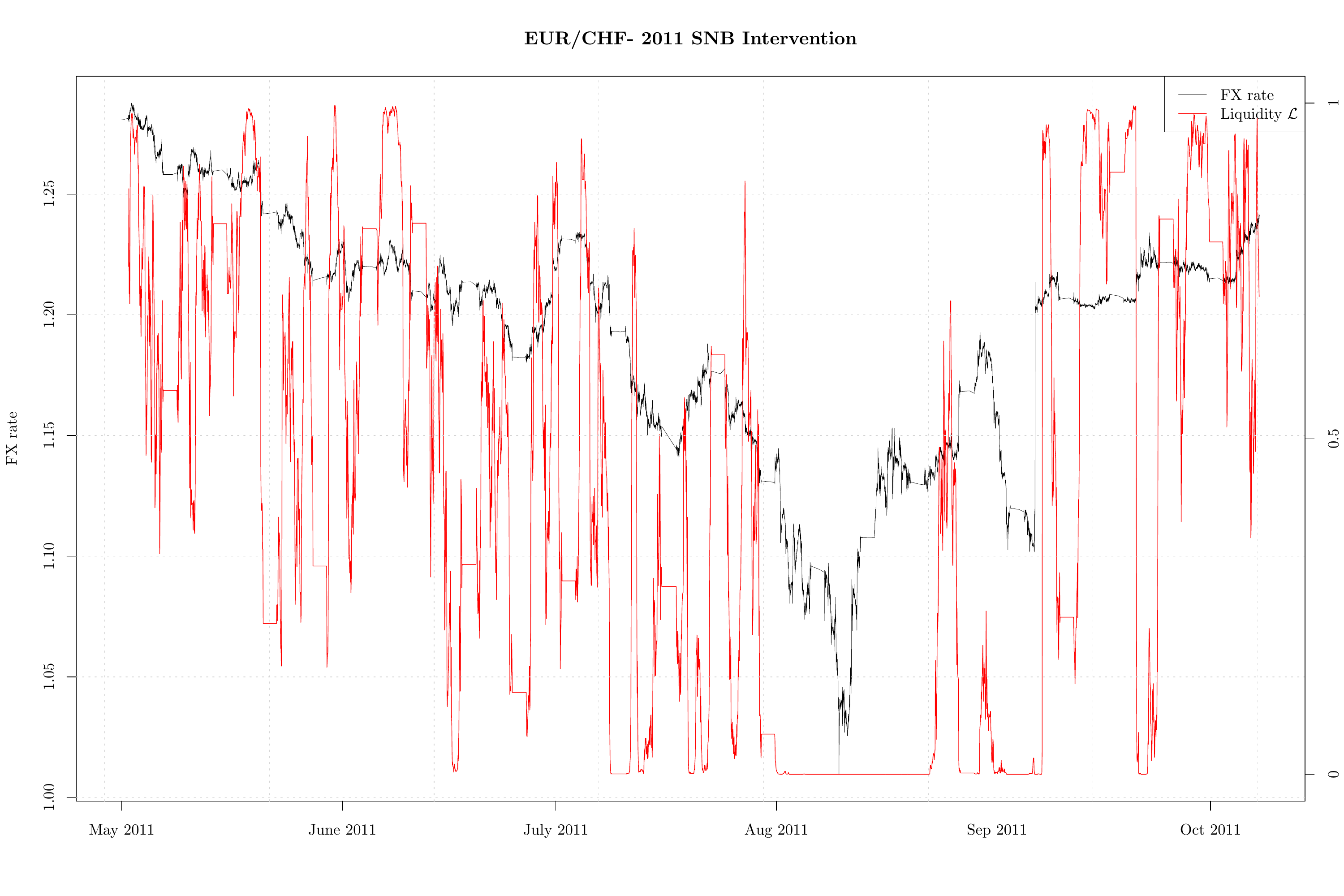}
\end{center}
\caption{
Time evolution of (left axis) tick-by-tick EUR/CHF exchange rate and
(right axis) the corresponding one minute Liquidity measurement
$\mathcal{L}$ in the months of Swiss National Bank intervention, setting
the floor of 1.20 on the EUR/CHF exchange rate.}
\label{fig:floor}
\end{figure}

Next we focus on the Swiss National Bank (SNB) setting the floor
rate after the Franc appreciated by a quarter of its value within a few
months as the debt crisis caused money to flee from the Euro zone. Since
then the SNB has systematically prevented the Euro from falling below
the 1.20 level, buying Euros and selling Francs, resulting in a massive
increase of FX reserves, which as of November 2013 stand at
approximately 434 billion Francs\footnote{Source: SNB
(http://www.snb.ch/ext/stats/imfsdds/pdf/deenfr/IMF.pdf)}\footnote{\textit{``Swiss to Join Euro: 73\% of Every Swiss Yearly Income Invested
in Euro via SNB''}, Dorgan, G., 2012 SNBCHF.com}. 

Figure~\ref{fig:floor} shows the time evolution of the tick-by-tick
EUR/CHF exchange rate and minute-by-minute Liquidity $\mathcal{L}$ in
the months of the SNB intervention. Our measurement shows slow but
steady deterioration of liquidity conditions during the time of Franc
appreciation. In fact, as the graph demonstrates the Liquidity
$\mathcal{L}$ measurement indicate a complete loss of liquidity during
the week proceeding spectacular near 1000pip (approximately $10\%$) gain
in Franc against Euro, reaching near parity on August 9th 2011. Our
measurement shows that illiquid market conditions continue in the next
weeks following a almost 20\% reversal, with the liquidity conditions
finally being restored after the SNB intervention on September 6th 2011
(Schmidt 2011).

\section{Conclusion}

In this article we use an event based framework to map continuous
financial data into the so-called intrinsic network. We define a method
for state contraction of intrinsic networks, and show that the network
has a consistent hierarchical structure, allowing multi-scale analysis
of financial data. We define the concept of Liquidity that characterises
the unlikeliness of price trajectories. Focusing on well know currency
crisis, we argue and show that the new metric has the ability to detect
and predict stress in financial markets. 

\section*{Acknowledgements}
The research leading to these results has received funding from the European Union Seventh Framework Programme (FP7/2007-2013) under grant agreement no 317534 (the Sophocles project).
\newpage
\begin{appendices}

\section{Intrinsic time framework}
\label{app:algo}

Formally, we map the time series of prices into sequences of directional
changes and price overshoots as follows. Let
$\Delta=\{\delta_0,\dots,\delta_{n_{\delta}-1}\}$ be the set of
$n_{\delta}$ directional change thresholds onto which time series is
mapped. The initial condition of $x_0$ the initial price; $t_0$, the
initial physical time; and $m_0$ the mode that switches between up and
down indicating in which direction the directional change is expected.
An initial condition affects at most the first two pairs (directional
change, overshoot), and let the subsequent pairs in the sequence to
synchronise with any other sequence obtained with a different
initialization the sequence obtained with a different initialisation.

A given $\delta_i$ discretises the time series into a set of prices
$X_i(t)=\{x^i_0(t_0^i),\dots, x_{n_i-1}^i(t_{n_i-1}^i), x(t)\}$
occurring at times $T_i(t)=\{t_0^i,\dots,t_{n_i-1}^i, t\}$ where
$x(t)=\big(\operatorname{bid}(t)+\operatorname{ask}(t)\big)/2$ is the
midprice at time $t$. We highlight that the last elements of the set
$\big(x(t),t\big)$ are temporary, as they do not correspond to a turning
point yet but represent the state of the process at time $t$. We compute
the number of turning points (i.e., the occurrence of a directional
change) as $n_e^i=\lfloor \frac{n_i}{2} \rfloor$. The series of
amplitude of directional changes $\Delta_i$ is defined as
\begin{equation}
\Delta_i(t)=\big\{\delta_0^i,\dots,\delta_{n_e^{i-1}}^i,\delta_{n_e^i}^i
(t)\big\}=\Big\{x_{2j+1}^i-x_{2j}^i\Big\} 
\label{eq:ap1}
\end{equation} 
where $0\leq j\leq n_e^i$. The discreteness of the time series of prices
prevents $|\delta_j^i|=\delta_i$. The discrepancy is, however, small and
is on average within the spread. The series of amplitudes of overshoots
$\Omega_i$ is written as 
\begin{equation}
\Omega_i(t)=\big\{\omega_0^i,\dots,\omega_{n_e^{i-1}-1}^i,\omega_{n_e^i}
^i(t)\big\}=\Big\{x_{2(j+1)}^i-x_{2j+1}^i\Big\}. 
\label{eq:ap2}
\end{equation} 

Duration of directional change or price overshoots are similarly defined
by replacing prices $x$ by physical time $t$ in equation (\ref{eq:ap1})
and (\ref{eq:ap2}). Algorithm $1$ shows a pseudocode that gives further
details on how to dissect the time series of prices.\newline

\noindent{\textbf{Algorithm 1. Dissect the price curve from time $t_0$  and measure overshoots with a $\delta_i$ price threshold}}\newline

\noindent{Require: initialise variables ($x^{ext}=x(t_0), mode$ is arbitrarily set to $up$, $X_i=x_0, T_i=t_0$)}\newline
1: update latest $X_i$ with $x(t)$\newline
2: update latest $T_i$ with $t$\newline
3: $~~~$\textbf{if} $mode$ is $down$ \textbf{then}\newline
4: $~~~~~~$   \textbf{if} $x(t)>x^{ext}$ \textbf{then}\newline
5: $~~~~~~~~~$       $x^{ext}\leftarrow x(t)$\newline
6: $~~~~~~$   \textbf{else if} $x(t)-x^{ext}\leq -\delta_i$ \textbf{then}\newline
7:  $~~~~~~~~~$      $x^{ext}\leftarrow x(t)$\newline
8:  $~~~~~~~~~$      $mode \leftarrow \text{up}$\newline
9:   $~~~~~~~~~$     $X_i\leftarrow x(t)$\newline
10:   $~~~~~~~~~$     $T_i\leftarrow t$\newline
11:  $~~~~~~$   \textbf{end if}\newline
12:  $~~~$\textbf{else if} $mode$ is $up$ \textbf{then}\newline
13:   $~~~~~~$  \textbf{if} $x(t)<x^{ext}$ \textbf{then}\newline
14:    $~~~~~~~~~$     $x^{ext}\leftarrow x(t)$\newline
15:     $~~~~~~$\textbf{else if} $x(t)-x^{ext}\geq \delta_i$ \textbf{then}\newline
16:      $~~~~~~~~~$  $x^{ext}\leftarrow x(t)$\newline
17:  $~~~~~~~~~$      $mode \leftarrow \text{down}$\newline
18:   $~~~~~~~~~$     $X_i\leftarrow x(t)$\newline
19:    $~~~~~~~~~$    $T_i\leftarrow t$\newline
20:   $~~~~~~$  \textbf{end if}\newline
21:  $~~~$\textbf{end if}

\section{The analytical Gaussian benchmark}
\label{app:FIT}

In the special case where the price follows a Brownian motion, the
transition matrix can be derived analytically. Since the hierarchical
nature of the intrinsic network allows deducing the transition matrix
for any number of thresholds by a contracting process, the problem
actually boils down to solving the two-thresholds case, which we will do
now. In case the transitions on the Intrinsic Network are modelled as first order Markov Chain, the matrix has the following form
\begin{equation}
W = \left(\begin{array}{cccc}
0 & 1 &  0 & 0 \\
1-\alpha & 0 &  0 & \alpha \\
\beta & 0 &  0 & 1-\beta \\
0 & 0 & 1 & 0 \\
\end{array}
\right)
\end{equation}
with the convention that the states are numbered $(0,0)=0$, $(1,0)=1$,
$(0,1)=2$ and $(1,1)=3$. We thus have only to calculate two
probabilities which will be taken as $\mathbb{P}\big((1,0) \to
(1,1)\big)=\alpha$ and $\mathbb{P}\big((0,1) \to (0,0)\big)=\beta$. For
reasons of convenience and without loss of generality, we will simplify
the calculations by considering the case where the thresholds are fixed
in terms of absolute value instead of a percentage, which makes little
difference -if any- as long the the volatility is not too hectic.\\

Let us now focus on the situation where the system just turned to the
$(1,0)$ state. As we can read it from $W$, it was previously in $(0,0)$
and just bounced back from some minimum by an amount $\delta_1$. Two
events may occur now

\begin{itemize}
	\item either the upward move goes further by an amount $\delta_2-\delta_1$ and the system turns to the $(1,1)$ state,
	\item either after having reached some maximum $M  < \delta_2-\delta_1$ the walk goes downward by an amount $\delta_1$ and the system turns back to $(0,0)$.
\end{itemize} 

The question is therefore to determine the probability of each of these
two scenarios to occur. This is somewhat reminiscent of the famous
gambler's ruin, but the situation is more involved here due to the
presence of two absorbing barriers, one of which is moving in time.

Let us denote $A\equiv x_0 + \Delta$ the upper fixed barrier and
$B\equiv M-\delta$ the moving lower one (for the sake of notation we put
$\Delta \equiv \delta_2-\delta_1$ and $\delta \equiv \delta_1$). We now
dissect the interval $(x_0,A)$ in small intervals $(x_0,x_0+\epsilon)$,
$(x_0+\epsilon,x_0+2\epsilon)$, ..., $(A-\epsilon,A)$ with $\epsilon
\equiv \Delta/n$ for some $n$. In order for the walk to reach $A$ before
$B$, it has, as a very first step, to reach reach $x_0+\epsilon$ before
$x_0 -\delta$ and then to reach $x_0+2\epsilon$ before
$x_0+\epsilon-\delta$, and so on. We are thus led to rewrite the
probability (let us denote it $\mathbb{P}(A\backslash B)$) to reach the
fixed threshold $A$ before the moving one $B$ as
\begin{equation}
\mathbb{P}(A\backslash B) = \prod_{k=1}^{\Delta/\epsilon} \mathbb{P}(x_0 + k \epsilon \backslash x_0 +(k-1)\epsilon - \delta \vert x_0 + (k-1) \epsilon \backslash x_0 + (k-2)\epsilon - \delta)
\end{equation}

But then invariance properties (Markovianity and translation invariance)
of Brownian motion allow us to simplify this expression as

\begin{equation}
\mathbb{P}(A\backslash B) = \big(\mathbb{P}(x_0 + \epsilon \backslash x_0 - \delta) \big)^{\Delta/\epsilon}
\end{equation}
and it remains to take the limit $\epsilon \to 0$.

Let us now simplify a bit further the notation and assume we have a
Brownian motion with mean $\mu$ and variance $\sigma^2$ starting at a
position $x_0$ somewhere between two absorbing barriers $U$ (upper) and
$L$ (lower). The probability density of finding the walk at position $x$
at time $t$ will obey the backward diffusion equation
\begin{equation}
\partial_t p(x_0,x,t) = \mu \partial_{x_0} p(x_0,x,t) + \frac{\sigma^2}{2} \partial_{x_0}^2 p(x_0,x,t)
\end{equation}
with boundary conditions $p(x_0,x,0) = \delta(x_0)$ and $p(x_0,U,t) =
p(x_0,L,t) = 0$. The best way to proceed is now to define
\begin{equation}
g(x_0,t) \equiv -\partial_t \int_{L}^{U} p(x_0,x,t) dx ,
\end{equation}
which denotes the probability to be absorbed around time $t$ by any of
the barriers, and then take the Laplace transform
\begin{equation}
G (x_0,s) \equiv \int_0^{\infty} e^{-st} g(x_0,t) dt
\end{equation}
which has the very interesting property that evaluating it at $s=0$
yields exactly the probability to be caught by any of the barriers. Some
standard manipulations allow us to transfer the backward equation to the
Laplace domain so as to obtain
\begin{equation}
s G(x_0,s) = \mu \partial_{x_0}G(x_0,s) + \frac{\sigma^2}{2} \partial_{x_0}^2 G(x_0,s)
\label{TBE}
\end{equation}
with boundary conditions $G(U,s) = G(L,s) =1$ (which means nothing but
immediate absorption if the walk starts on either barrier).

We then split the total probability of absorption as $g_{-}(x_0,t) +
g_{+}(x_0,t)$, where $g_{\pm}(x_0,t)$ denotes the probability of
absorption by the upper, respectively lower, barrier. The transform is
split accordingly as $G_{-}(x_0,s) + G_{+}(x_0,s)$ with boundary
conditions $G_{+}(U,s) = G_{-}(L,s) = 1$ and $G_{+}(L,s) = G_{-}(U,s) =
0$. Equation (\ref{TBE}) can thus be solved separately for $G_{+}$ and
$G_{-}$. We use the standard ansatz $G_{\pm} = \exp (\theta x_0)$ which
boils down the differential equation to a quadratic algebraic equation
for $\theta$ easily solved to yield
\begin{equation}
\theta_{1,2} = \frac{-\mu \mp \sqrt{\mu^2 + 2s\sigma^2}}{\sigma^2}
\end{equation}
(\ref{TBE}) is then solved by
\begin{equation}
G_{\pm}(x_0,s) = K_{1} e^{\theta_{1} x_0} + K_{2} e^{\theta_{2} x_0}
\end{equation}
for constants chosen to match the boundary conditions. We skip the
details to quote the expression for $G_{+}$, taking according to our
previous notations $U=x_0 + \epsilon$ and $L=x_0 - \delta$, and putting
$s=0$,
\begin{equation}
G_{+}(x_0,0) = \frac{1- \exp \left( \frac{-2\delta \vert \mu \vert}{\sigma^2} \right)}{1- \exp \left( \frac{-2(\delta+\epsilon) \vert \mu \vert}{\sigma^2} \right)} \exp \left(  \frac{\epsilon (\mu - \vert \mu \vert)}{\sigma^2} \right)
\end{equation}

This quantity is therefore the probability to get caught by the upper
barrier without having ever met the lower one, which is $P(x_0+\epsilon
\backslash x_0-\delta)$ we introduced at the beginning. It thus remains
to calculate
\begin{equation}
\lim_{\epsilon \to 0} G_{+} (x_0,0) ^{\Delta/ \epsilon}
\end{equation}
which is easily found to be
\begin{equation}
\mathbb{P}(A \backslash B) = \exp \left( -\frac{\Delta}{\sigma^2} \cdot \frac{(\vert \mu \vert - \mu) + (\vert \mu \vert + \mu) \exp \left( \frac{-2\delta \vert \mu \vert}{\sigma^2} \right)}{1-\exp \left( \frac{-2\delta \vert \mu \vert}{\sigma^2} \right)} \right)
\end{equation}

This expression happens to simplify in the driftless case to the
harmless formula
\begin{equation}
\mathbb{P}(A \backslash B) = \exp \left( -\frac{\Delta}{\delta} \right).
\end{equation}

This is the expression we were searching for the probability of
transitioning from $(1,0)$ to $(1,1)$. The very same reasoning applies
using $G_{-}$ for the transition from $(0,1)$ to $(0,0)$, while other
transitions are now trivial. $W$ for a two-thresholds system can
therefore be written as
\begin{equation}
W = \left(\begin{array}{cccc}
0 & 1 &  0 & 0 \\
1-\exp \left(-\frac{\delta_2-\delta_1}{\delta_1} \right) & 0 &  0 &  \exp \left(- \frac{\delta_2-\delta_1}{\delta_1} \right) \\
\exp \left(-\frac{\delta_2-\delta_1}{\delta_1} \right) & 0 &  0 & 1-\exp \left(- \frac{\delta_2-\delta_1}{\delta_1} \right) \\
0 & 0 & 1 & 0 \\
\end{array}
\right)
\end{equation}

Obviously in that case the ratio of the thresholds only matters, and not
the thresholds themselves.

The derivation established that the probability of overshoot
$\omega(\delta_1;\sigma)$ reaching the length $\delta_2-\delta_1$ equals
$\exp \left(-\frac{\delta_2-\delta_1}{\delta_1} \right)$, i.e.
\begin{equation}
\mathbb{P}(\omega(\delta_1;\sigma)\geq \delta_2-\delta_1)=\exp \left(-\frac{\delta_2-\delta_1}{\delta_1} \right)
\end{equation}
hence we conclude that the overshoot lengths are exponentially
distributed. From there it straightforwardly follows that the average
length of the overshoot equals the directional change threshold
\begin{equation}
\mathbb{E}[\omega(\delta;\sigma)]=\delta
\end{equation}
thus proving the fundamental intrinsic theorem ~\ref{th:fit}.

\section{Contraction of probabilities}
\label{app:contrProb}
We demonstrate that assuming the transitions on the Intrinsic Network are modelled as a first order Markov Chain, there is an explicit connection between the transition matrix $W$ of the $n$-dimensional Intrinsic Network and transition matrix $\widehat{W}$ of the $n-1$ dimensional Intrinsic network obtained through the contraction process.

Let us assume that the current state of the network is $\mathcal{I}_k^{(i)}$ and we are interested in finding the probability of transitioning to state $\mathcal{I}_j^{(i)}$. Note that Island $\mathcal{I}_k^{(i)}$ consists of states $\{\mathcal{I}_{2k}^{(i-1)}, \mathcal{I}_{2k+1}^{(i)}\}$, while Island $\mathcal{I}_j^{(i)}$ consists of state $\{\mathcal{I}_{2j}^{(i-1)}, \mathcal{I}_{2j+1}^{(i)}\}$.  For $\operatorname{mod}(k+j,2)=0$, from $\mathcal{I}_{2k+1}^{(i-1)}$ the system can directly transition to $\mathcal{I}_{2j+1}^{(i-1)}$ with probability $\mathbb{P}(\mathcal{I}^{(i-1)}_{2k+1}\rightarrow \mathcal{I}^{(i-1)}_{2j+1})$. On the other hand, the system can oscillate once within Island $\mathcal{I}_k^{(i)}$ before proceeding to Island $\mathcal{I}_j^{(i)}$, i.e. transition from $\mathcal{I}_{2k+1}^{(i-1)}$ to $\mathcal{I}_{2k}^{(i-1)}$ and back to $\mathcal{I}_{2k+1}^{(i-1)}$, before transitioning to $\mathcal{I}_{2j+1}^{(i)}$, with probability 

$$\mathbb{P}(\mathcal{I}^{(i-1)}_{2k+1}\rightarrow \mathcal{I}^{(i-1)}_{2k})\cdot \mathbb{P}(\mathcal{I}^{(i-1)}_{2k}\rightarrow \mathcal{I}^{(i-1)}_{2k+1})\cdot\mathbb{P}(\mathcal{I}^{(i-1)}_{2k+1}\rightarrow \mathcal{I}^{(i-1)}_{2j+1}).$$ Likewise, oscillation within Island $\mathcal{I}_k^{(i)}$ can occur $k$ times, before proceeding to Island $\mathcal{I}_{j}^{(i)}$, with probability, 
$$\big(\mathbb{P}(\mathcal{I}^{(i-1)}_{2k}\rightarrow \mathcal{I}^{(i-1)}_{2k+1})\cdot \mathbb{P}(\mathcal{I}^{(i-1)}_{2k+1}\rightarrow \mathcal{I}^{(i-1)}_{2k})\big)^k\cdot \mathbb{P}(\mathcal{I}^{(i-1)}_{2k+1}\rightarrow \mathcal{I}^{(i-1)}_{2j+1}),$$
hence the probability to transition from Island $\mathcal{I}_k^{(i)}$ before to Island $\mathcal{I}_j^{(i)}$ equals
\begin{align}
\mathbb{P}(\mathcal{I}_k^{(i)}\rightarrow \mathcal{I}_j^{(i)})&=\sum_{k=0}^{\infty}\big(\mathbb{P}(\mathcal{I}^{(i-1)}_{2k}\rightarrow \mathcal{I}^{(i-1)}_{2k+1})\cdot \mathbb{P}(\mathcal{I}^{(i-1)}_{2k+1}\rightarrow \mathcal{I}^{(i-1)}_{2k})\big)^k\cdot \mathbb{P}(\mathcal{I}^{(i-1)}_{2k+1}\rightarrow \mathcal{I}^{(i-1)}_{2j+1})\\
&=\frac{\mathbb{P}(\mathcal{I}^{(i-1)}_{2k+1}\rightarrow \mathcal{I}^{(i-1)}_{2j+1})}{1-\mathbb{P}(\mathcal{I}^{(i-1)}_{2k}\rightarrow \mathcal{I}^{(i-1)}_{2k+1})\cdot \mathbb{P}(\mathcal{I}^{(i-1)}_{2k+1}\rightarrow \mathcal{I}^{(i-1)}_{2k})}.
\end{align}
For $\operatorname{mod}(k+j,2)=1$ and $k>j$, the system has to make an interim transition from $\mathcal{I}^{(i-1)}_{2k+1}$ to $\mathcal{I}^{(i-1)}_{2k}$, before transitioning from Island $\mathcal{I}_k^{(i)}$ to Island $\mathcal{I}_{j}^{(i)}$, with probability
$$\mathbb{P}(\mathcal{I}^{(i-1)}_{2k+1}\rightarrow \mathcal{I}^{(i-1)}_{2k})\cdot \mathbb{P}(\mathcal{I}^{(i-1)}_{2k}\rightarrow \mathcal{I}^{(i-1)}_{2j}).$$
Likewise, as before the system can oscillate $k$ times within Island $\mathcal{I}_k^{(i)}$ before proceeding to Island $\mathcal{I}_{j}^{(i)}$, with probability 
$$\mathbb{P}(\mathcal{I}^{(i-1)}_{2k+1}\rightarrow \mathcal{I}^{(i-1)}_{2k})\cdot \big(\mathbb{P}(\mathcal{I}^{(i-1)}_{2k}\rightarrow \mathcal{I}^{(i-1)}_{2k+1})\cdot \mathbb{P}(\mathcal{I}^{(i-1)}_{2k+1}\rightarrow \mathcal{I}^{(i-1)}_{2k})\big)^k\cdot \mathbb{P}(\mathcal{I}^{(i-1)}_{2k}\rightarrow \mathcal{I}^{(i-1)}_{2j}),$$
hence the probability to transition from Island $\mathcal{I}_k^{(i)}$ before to Island $\mathcal{I}_j^{(i)}$ equals
\begin{align}
\mathbb{P}(\mathcal{I}_k^{(i)}\rightarrow \mathcal{I}_j^{(i)})&=
\sum_{k=0}^{\infty}\mathbb{P}(\mathcal{I}^{(i-1)}_{2k+1}\rightarrow \mathcal{I}^{(i-1)}_{2k})\cdot \big(\mathbb{P}(\mathcal{I}^{(i-1)}_{2k}\rightarrow \mathcal{I}^{(i-1)}_{2k+1})\cdot \mathbb{P}(\mathcal{I}^{(i-1)}_{2k+1}\rightarrow \mathcal{I}^{(i-1)}_{2k})\big)^k\cdot \mathbb{P}(\mathcal{I}^{(i-1)}_{2k}\rightarrow \mathcal{I}^{(i-1)}_{2j})\\
&=\frac{\mathbb{P}(\mathcal{I}^{(i-1)}_{2k+1}\rightarrow \mathcal{I}^{(i-1)}_{2k})\cdot \mathbb{P}(\mathcal{I}^{(i-1)}_{2k}\rightarrow \mathcal{I}^{(i-1)}_{2j})}{1-\mathbb{P}(\mathcal{I}^{(i-1)}_{2k}\rightarrow \mathcal{I}^{(i-1)}_{2k+1})\cdot \mathbb{P}(\mathcal{I}^{(i-1)}_{2k+1}\rightarrow \mathcal{I}^{(i-1)}_{2k})}.
\end{align}
Similarly, it can be shown that for $\operatorname{mod}(k+j,2)=1$ and $k<j$, it can be shown
\begin{align}
\mathbb{P}(\mathcal{I}_k^{(i)}\rightarrow \mathcal{I}_j^{(i)})&=
\sum_{k=0}^{\infty}\mathbb{P}(\mathcal{I}^{(i-1)}_{2k}\rightarrow \mathcal{I}^{(i-1)}_{2k+1})\cdot \big(\mathbb{P}(\mathcal{I}^{(i-1)}_{2k}\rightarrow \mathcal{I}^{(i-1)}_{2k+1})\cdot \mathbb{P}(\mathcal{I}^{(i-1)}_{2k+1}\rightarrow \mathcal{I}^{(i-1)}_{2k})\big)^k\cdot \mathbb{P}(\mathcal{I}^{(i-1)}_{2k+1}\rightarrow \mathcal{I}^{(i-1)}_{2j+1})\\
&=\frac{\mathbb{P}(\mathcal{I}^{(i-1)}_{2k}\rightarrow \mathcal{I}^{(i-1)}_{2k+1})\cdot \mathbb{P}(\mathcal{I}^{(i-1)}_{2k+1}\rightarrow \mathcal{I}^{(i-1)}_{2j+1})}{1-\mathbb{P}(\mathcal{I}^{(i-1)}_{2k}\rightarrow \mathcal{I}^{(i-1)}_{2k+1})\cdot \mathbb{P}(\mathcal{I}^{(i-1)}_{2k+1}\rightarrow \mathcal{I}^{(i-1)}_{2k})}.
\end{align}

\section{Transition probability derivation}

We will demonstrate the derivation of the analytic expressions of
transition probabilities presented in Section 5. Firstly, we prove the
claim holds for $3$-dimensional intrinsic network, let
$\delta_1<\delta_2<\delta_3$ denote the ordered directional change
thresholds, $\mathcal{IN}(3;\{\delta_1,\delta_2,\delta_3\};W)$
3-dimensional intrinsic network and
$\mathcal{IN}(2;\{\delta_2,\delta_3\};\widehat{W})$ the contracted
intrinsic network. Since the contracted intrinsic network is
2-dimensional, it is know that \[\mathbb{P}\big((1,0)\rightarrow
(1,1)\big)=e^{-\frac{\delta_3-\delta_2}{\delta_2}},\] while the explicit
analytic expression of transition probabilities between 3-  and
contracted 2-dimensional intrinsic network presented in Section 4 states
that
\[e^{-\frac{\delta_3-\delta_2}{\delta_2}}=\frac{\mathbb{P}\big((1,1,0)
\rightarrow (1,1,1)\big)}{1-\Big(1-\mathbb{P}\big((1,1,0)\rightarrow
(1,1,1)\big)\Big)\Big(1-e^{-\frac{\delta_2-\delta_1}{\delta_1}}\Big)}\]
and untangling the formula we find
\begin{equation}
\mathbb{P}\big((1,1,0)\rightarrow (1,1,1)\big)=\frac{e^{-\frac{\delta_3-\delta_2}{\delta_2}}e^{-\frac{\delta_2-\delta_1}{\delta_1}}}{1-e^{-\frac{\delta_3-\delta_2}{\delta_2}}\big(1-e^{-\frac{\delta_2-\delta_1}{\delta_1}}\big)}
\end{equation}
yielding the desired expression. Let us assume that the claim holds for
$n$-dimensional intrinsic network, and we will prove that the claim
holds for $n+1$. Let $\delta_1<\dots<\delta_{n+1}$ denote the ordered
directional change thresholds,
$\mathcal{IN}(n+1;\{\delta_1,\dots,\delta_{n+1}\};W)$ $n+1$-dimensional
intrinsic network and
$\mathcal{IN}(n;\{\delta_2,\dots,\delta_{n+1}\};\widehat{W})$ the
contracted intrinsic network. Since the contracted intrinsic network is
$n$-dimensional, together with explicit analytic expression of
transition probabilities between $n+1$-  and contracted $n$-dimensional
intrinsic network presented in Section 4 states that
\begin{align*}
\frac{\prod_{k=3}^{n+1} e^{-\frac{\delta_k-\delta_{k-1}}{\delta_{k-1}}}}{1-\sum_{k=3}^{n}\big(1-e^{-\frac{\delta_k-\delta_{k-1}}{\delta_{k-1}}}\big)\prod_{j=k+1}^{n+1} e^{-\frac{\delta_j-\delta_{j-1}}{\delta_{j-1}}}}=\frac{\mathbb{P}\big((1,\dots,1,0)\rightarrow (1,\dots,1,1)\big)}{1-\Big(1-\mathbb{P}\big((1,\dots,1,0)\rightarrow (1,\dots,1,1)\big)\Big)\Big(1-e^{-\frac{\delta_2-\delta_1}{\delta_1}}\Big)}
\end{align*}
we find
\begin{align}
\mathbb{P}\big((1,\dots,1,0)\rightarrow~&(1,\dots,1,1) \big) =\\ &\frac{\prod_{k=3}^{n+1} e^{-\frac{\delta_k-\delta_{k-1}}{\delta_{k-1}}}\cdot e^{-\frac{\delta_2-\delta_1}{\delta_1}}}{1-\sum_{k=3}^{n}\big(1-e^{-\frac{\delta_k-\delta_{k-1}}{\delta_{k-1}}}\big)\prod_{j=k+1}^{n+1} e^{-\frac{\delta_j-\delta_{j-1}}{\delta_{j-1}}}-(1-e^{-\frac{\delta_2-\delta_1}{\delta_1}})\cdot\prod_{k=3}^{n+1}e^{-\frac{\delta_{k}-\delta_{k-1}}{\delta_{k-1}}}}
\end{align}
obtaining the desired formula.

\section{Convergence theorems}
\label{app:convThe}

In this section we present Shannon-McMillan-Brieman theorem on
convergence of sample entropy and the corresponding central limit
theorem, in a more general setting following (Pfister \textit{et. al.}
2001); let $\{X_t\}_{t\geq 1}$ denote an ergodic finite state process
which gives rise to the conditional probability sequence $\{A_t\}_{t\geq
1}$ where $A_t=\mathbb{P}(X_t|\mathbf{X}_1^{t-1})$ and $H(\mathcal{X})$
denotes the entropy of the process, while $\widehat{H}_n(\mathcal{X})$
denotes the sample entropy rate, i.e.
\begin{equation}
\widehat{H}_n(\mathcal{X})=-\frac{1}{n}\mathbb{P}(\mathbf{X}_1^n).
\end{equation}
\begin{tm}{(Shannon-McMillan-Breiman theorem)}
\begin{equation}
\widehat{H}_n(\mathcal{X})=-\frac{1}{n}\sum_{t=1}^n\operatorname{log}\mathbb{P}(X_t|\mathbf{X}_1^{t-1})\rightarrow H(\mathcal{X})~~ (a.s.)
\end{equation}
\end{tm}
\begin{tm}{(Central limit theorem)} 
If  $$\lim_{t\rightarrow
\infty}\mathbb{E}[(-\operatorname{log}A_t)^{2+\varepsilon}]<\infty,$$
then the sample entropy rate obeys a central limit theorem of the form
\begin{equation}
\sqrt{n}\big[\widehat{H}_n(\mathcal{X})-H(\mathcal{X})\big]\rightarrow N(0,\sigma^2).
\end{equation}

The variance, $\sigma^2$, of the estimate is given by
\begin{equation}
\sigma^2=R(0)+2\sum_{\tau=1}^{\infty}R(\tau)
\end{equation}
where $R(\tau)=\lim_{t\rightarrow \infty}\mathbb{E}[(-\operatorname{log}A_t-H(\mathcal{X}))(-\operatorname{log}(A_{t-\tau})-H(\mathcal{X}))]$. If we also have that 
$$
\lim_{t\rightarrow \infty}\mathbb{E}[(-\operatorname{log}A_t)^{4+\varepsilon}]<\infty
$$
then we can estimate the variance using finite truncations of (22) with $R(\tau)$ set to the sample autocorrelation
\begin{equation}
\widehat{R}_n(\tau)=\frac{1}{n-\tau}\sum_{t=\tau+1}^n \big(-\operatorname{log}A_t-\widehat{H}_n(\mathcal{X})\big)
\big(-\operatorname{log}A_{t-\tau}-\widehat{H}_n(\mathcal{X})\big).
\end{equation}
\end{tm}

\end{appendices}

\end{document}